\documentclass[12pt]{article}
\usepackage{amsmath,amssymb,array,calc,rotating,epsfig,psfrag,amscd, cite}
\usepackage{epsfig}
\usepackage{amsmath}
\makeatletter
\renewcommand\section{\@startsection {section}{1}{\z@}%
                                   {-3.5ex \@plus -1ex \@minus -.2ex}%nn
                                   {2.3ex \@plus.2ex}%
                                   {\normalfont\large\bfseries}}
\renewcommand\subsection{\@startsection{subsection}{2}{\z@}%
                                     {-3.25ex\@plus -1ex \@minus -.2ex}%
                                     {1.5ex \@plus .2ex}%
                                     {\normalfont\bfseries}}
\makeatother

\let\non\nonumber

\let\k=\kappa

\let\S=\Sigma

\newcommand{\bea}{\begin{eqnarray}}
\newcommand{\eea}{\end{eqnarray}}
\newcommand{\be}{\begin{equation}}
\newcommand{\ee}{\end{equation}}

\newcommand{\pref}[1]{(\ref{#1})}

\def\coeff#1#2{{\textstyle\frac{#1}{#2}}}
\newcommand{\hlf}{\frac{1}{2}}

\newcommand{\Z}{{\mathbb Z}}
\newcommand{\Rr}{{\mathbb R}}
\newcommand{\Cc}{{\mathbb C}}
\newcommand{\Ss}{{\mathbb S}}
\newcommand{\Tt}{{\mathbb T}}

\newcommand{\SO}{\operatorname{SO}}

\newcommand{\m}{\mu}

\newcommand{\p}{\partial}

\newcommand{\ap}{\alpha'}
\newcommand{\al}{\alpha}

\newcommand{\gym}{g_{\rm YM}}

\newcommand{\lstr}{\ell_s}
\newcommand{\lpl}{\ell_p}

\newcommand{\C}[1]{$(\ref{#1})$}

\def\ie{{\it i.e.}}

\def\IZ{\relax\ifmmode\mathchoice
{\hbox{\cmss Z\kern-.4em Z}}{\hbox{\cmss Z\kern-.4em Z}}
{\lower.9pt\hbox{\cmsss Z\kern-.4em Z}} {\lower1.2pt\hbox{\cmsss
Z\kern-.4em Z}}\else{\cmss Z\kern-.4em Z}\fi}
\def\IR{\relax{\rm I\kern-.18em R}}

\def\one{{\hbox{ 1\kern-.8mm l}}}

\def\R{{\rm R}}

\def\nn{{\bf n}}

\def\Tr{{\rm Tr\,}}

\newlength{\bredde}
\def\slash#1{\settowidth{\bredde}{$#1$}\ifmmode\,\raisebox{.15ex}{/}
\hspace*{-\bredde} #1\else$\,\raisebox{.15ex}{/}\hspace*{-\bredde}
#1$\fi}

\newsavebox{\zzzbar}
\sbox{\zzzbar}
  {\setlength{\unitlength}{0.9em}
  \begin{picture}(0.6,0.7)
  \thinlines
  \put(0,0){\line(1,0){0.6}}
  \put(0,0.75){\line(1,0){0.575}}
  \multiput(0,0)(0.0125,0.025){30}{\rule{0.3pt}{0.3pt}}
  \multiput(0.2,0)(0.0125,0.025){30}{\rule{0.3pt}{0.3pt}}
  \put(0,0.75){\line(0,-1){0.15}}
  \put(0.015,0.75){\line(0,-1){0.1}}
  \put(0.03,0.75){\line(0,-1){0.075}}
  \put(0.045,0.75){\line(0,-1){0.05}}
  \put(0.05,0.75){\line(0,-1){0.025}}
  \put(0.6,0){\line(0,1){0.15}}
  \put(0.585,0){\line(0,1){0.1}}
  \put(0.57,0){\line(0,1){0.075}}
  \put(0.555,0){\line(0,1){0.05}}
  \put(0.55,0){\line(0,1){0.025}}
  \end{picture}}

\newcommand{\ena}{\end{eqnarray}}
\newcommand{\beqa}{\begin{eqnarray}}
\newcommand{\eeqa}{\end{eqnarray}}

\def\FF{{\cal F}}

\def\k{\kappa}

\def\m{\mu}

\def\S{\Sigma}

\renewcommand{\SO}{\operatorname{SO}}

\newcommand{\Dslash}{\ensuremath \raisebox{0.025cm}{\slash}\hspace{-0.32cm} D}

\newcommand{\lp}{\left(}
\newcommand{\rp}{\right)}
\newcommand{\ls}{\left[}
\newcommand{\rs}{\right]}
\newcommand{\LR}{R}
\def\xx{{\bf x}}

\def\ym{{\rm YM}}
\def\One{{1\hskip -3pt {\rm l}}}

\begin{document}
\begin{titlepage}

\begin{center}

{April 21, 2009} \hfill EFI-09-13,  UTTG-02-09
%\today
%\hfill         \phantom{xxx}

\vskip 2 cm
{\Large \bf Non-Supersymmetric String Theory}\\
\vskip 1.25 cm { Emil J. Martinec\footnote{email address:
ejmartin@uchicago.edu}$^{a}$, Daniel Robbins\footnote{email address:
robbins@zippy.ph.utexas.edu}$^{b}$ and Savdeep Sethi\footnote{email
address:
 sethi@uchicago.edu}}$^{a}$\\
{\vskip 0.5cm $^{a}$ Enrico Fermi Institute, University of Chicago,
Chicago, IL
60637, USA\\}

{\vskip 0.5cm $^{b}$ Physics Department, University of Texas at Austin, Austin, TX 78712\\}

\end{center}
\vskip 2 cm

\begin{abstract}
\baselineskip=18pt

A class of non-supersymmetric string backgrounds can be constructed
using twists that involve space-time fermion parity.
%Some of these backgrounds include bulk closed string tachyons.
We propose a non-perturbative definition of string theory in these
backgrounds via gauge theories
%We propose a non-perturbative definition of string theory in a class
%of backgrounds with supersymmetry broken by twists that involve
%space-time fermion parity. The dual descriptions are gauge theories
with supersymmetry softly broken by twisted boundary conditions. The
perturbative string spectrum is reproduced, and qualitative effects
of the interactions are discussed. 
Along the way, we find an interesting mechanism for inflation. 
The end state of closed string
tachyon condensation is a highly excited state in the gauge theory
which, in all likelihood, does not have a geometric interpretation.

\end{abstract}

\end{titlepage}

\pagestyle{plain}
%\baselineskip=18pt
% Try a wider skip
\baselineskip=19pt
%%%%%%%%%%%%%%%%%%%%%%%%%%%%%%%%%%%%%%%%%%%%%%%%%%%%%%%%%%%%%%%%%%%%%%%%%%%%%%
\section{Introduction}

The description of string theory in non-supersymmetric backgrounds
remains a poorly understood yet critical question. The description
of such backgrounds is typically given in terms of an approximate
semi-classical solution of string theory. This can be a solution of
the supergravity equations of motion or a world-sheet conformal
field theory. This description is typically static to leading order
in the string coupling expansion. However, quantum corrections to
this leading order picture can dramatically alter the physics
largely because the background becomes dynamical with a non-trivial
cosmology.

Our aim here is to investigate one of the simplest classes of string
backgrounds that break supersymmetry -- namely, the
 compactification studied by by
 Rohm~\cite{Rohm:1983aq}, and by Scherk and Schwarz~\cite{Scherk:1978ta},
  where the breaking
  is introduced by a twisted boundary condition along a cycle of a
  torus.\footnote{Variant backgrounds of this sort can be naively
  related by T-duality along the twisted circle, which is a good perturbative
   symmetry of string theory on certain static space-times.}
   A Matrix theory dual appears to exist non-perturbatively in a sense
   similar to the Matrix Big Bang picture of a cosmological singularity.
    Namely, there is a period of time in which a conventional notion of
    space-time with gravitons makes sense. However, in the matrix theory dual,
    space-time eventually ``dissolves'' with graviton dynamics replaced by gauge
     dynamics. This behavior has been seen in light-like dilaton backgrounds
      with a null singularity in the far past but it appears to be a more general
      picture for non-supersymmetric string backgrounds~%
\cite{Craps:2005wd, Li:2005ti, Robbins:2005ua, Chu:2006pa,
Das:2006dr, Das:2006dz, Lin:2006ie, Li:2005sz, She:2005mt,
Chen:2005bk, Craps:2006xq, Martinec:2006ak}.

The most straightforward construction of a dual background was
studied by Banks and Motl~\cite{Banks:1999tr}, who argued that the
non-perturbative description of the twisted compactification
(obtained by the standard Matrix theory prescription) fails to make
sense as a theory of Minkowski space-time with asymptotic gravitons
for any period of time.  The difficulty lies in a property of the
standard Matrix theory orbifold construction; namely, that it leads
to a gauge theory that is not asymptotically supersymmetric in the
UV.

We will propose an alternative approach to defining this background
which leads to a strongly coupled non-supersymmetric gauge theory.
The key difference is that in the present proposal, the
supersymmetry breaking is a soft breaking by boundary conditions,
allowing the delicate cancellations required for a dual
gravitational interpretation of the gauge theory.  The boundary
condition is twisted by an element of the R-symmetry, of the sort
studied in~\cite{Bergman:2000cw, Bergman:2001rw,Ganguli:2003kr}. We
will focus on R-symmetry twists that preserve no supersymmetry.
Similar boundary conditions have also been considered in the context
of supergravity Melvin solutions for which there is a vast
literature; see, for example~\cite{Costa:2000nw}.

Backgrounds of the kind we will describe provide nice laboratories
for studying closed string tachyon condensation.  Tachyon
condensation is just another name for a scalar field rolling along
an unstable direction in its potential, and is thus of interest in
many contexts, not the least of which is inflation.
%Defining string theories with bulk closed string tachyons is,
%however, quite mysterious. For example, the definition of physical
%observables depends on understanding the end point of tachyon
%condensation.
Tachyon condensation in string theory has been studied in the
context of both open strings~\cite{Sen:1998sm}\ and localized
instabilities of closed strings~\cite{Adams:2001sv,Harvey:2001wm}.
It has also been suggested that perturbative tachyon condensation
might play a role in the resolution of cosmological
singularities~\cite{Horowitz:2005vp,Horowitz:2006mr}. By tuning
parameters, it may be possible to find regimes in the
non-perturbative formulation of the twisted compactification where
perturbative tachyons condense before world-sheet string theory
becomes invalid. It would be interesting to see whether this can
actually happen in the context of these complete non-perturbative
models.

There will also be many related backgrounds in which space-time
physics is described non-perturbatively by gauge theories of
different dimension, matter content and dynamics.  Understanding the
initial and final states of these theories is a crucial issue in
determining how to define observables. The physical phenomena to be
found in string cosmology are likely to be at least as rich as the
phenomena seen in gauge theory.

A very natural class of models to study are the ten-dimensional
non-supersymmetric strings. Among this class of models, we will
mainly consider type 0 string theory and the tachyon free heterotic
model described in~\cite{Dixon:1986iz}. The latter model is
particularly interesting because the one-loop energy density is
positive. There are various conjectures about dualities and the end
points of tachyon condensation in these
models~\cite{Bergman:1999km,Blum:1997cs, Blum:1997gw}. It would be
interesting to examine these conjectures in the context of a
non-perturbative definition.

%A byproduct of our construction is yet another mechanism for
%inflation.  
Our construction also suggests another possible mechanism for inflation in string theory.
The gauge theory dual to the tachyon condensation
process is the unwinding of a state of large winding holonomy.  The
holonomy is prevented from unravelling quickly at strong coupling,
leading to an extended period of time where the state of the system
is ordered with energy well above the ground state.  
In these simple models, the space-time picture of the perturbative instability does not admit slow-roll inflation, but looking for regimes in which slow-roll could proceed is a promising direction for future research. This would place inflation (which is sensitive to Planck scale physics) in an ultraviolet complete framework.

This paper is organized as follows: in section  two, we describe the
backgrounds of interest and review their perturbative string
spectrum. We examine aspects of the cosmology associated to the
one-loop corrected equations of motion. In section three, we turn to
the non-perturbative description obtained using Matrix
theory~\cite{Banks:1996vh} (for reviews,
see~\cite{Banks:1999az,Taylor:2001vb}) and present a discussion of
the dynamics of this model in section four.  Section five presents a discussion and comparison to previous work.  A computation of the one-loop potential in the dual gauge theory is relegated to an appendix.

\section{The Perturbative String Analysis}
\label{perturbativestuff}

One of the simplest ways to break supersymmetry in string theory is
to consider a circle compactification. The presence of a non-trivial
loop requires a choice of  spin structure for space-time fermions.
If we choose anti-periodic boundary conditions for some fermions, we
mildly break any supersymmetry preserved by the uncompactified
background.

In string theory, such backgrounds  usually develop closed string
tachyon instabilities if the radius of the circle approaches string
scale. In this section, we will describe some aspects of
perturbative string theory on these kinds of backgrounds. There are
at least two tunable parameters at early times: the string coupling
and the radius of the circle which controls the scale of
supersymmetry breaking. The effects of the supersymmetry breaking
can therefore be made small for a long time and we expect a good
gravitational description over that period of time.

We will also describe the T-dual compactification. At the level of
world-sheet conformal field theory, the physics of these backgrounds
is identical. However, T-duality is only a symmetry of perturbative
string theory. It is quite possible, for example, that a twisted
compactification of type IIA string theory might have a very
different non-perturbative definition from the T-dual type IIB
background. Indeed, the T-dual description need not even exist. We
will return to this point in section~\ref{T-dual}.

\subsection{Closed strings on the Rohm circle}

We will start by analyzing perturbative closed strings  on a circle
of radius $L$ with anti-periodic boundary conditions for space-time
fermions. Much of our notation follows~\cite{Polchinski:1998rr}. We
construct this theory by taking either type IIA or IIB in
ten-dimensional flat space-time and quotienting by the group
generated by a $2\pi L$ translation in one direction (which we will
label $Y$) combined with multiplication by $(-1)^{F_s}$, where $F_s$
is the space-time fermion number.

In the untwisted sector, we can compute the partition sum trace with
an insertion of $g^n$, where $g$ is the generator of our group. Letting
$k$ be the momentum in the $Y$ direction and $p$ be the momenta in
the other directions, we find \bea
{\mathcal Z}_0^n & =  & \Tr_{\mathrm{untwisted}}e^{2\pi iLkn}\lp q\bar q\rp^{\frac{\al'}{4}(k^2+p^2)}\ls \frac{1+e^{\pi iF}}{2}q^{N_L-\frac{1}{2}}-e^{\pi in}\frac{1+e^{\pi iF}}{2}q^{N_L}\rs\non\\
& & \times\ls\frac{1+e^{\pi i\tilde F}}{2}\bar q^{N_R-\frac{1}{2}}-e^{\pi in}\frac{1\pm e^{\pi i\tilde F}}{2}\bar q^{N_R}\rs, \non\\
& = & iV_{10}\int\frac{d^9pdk}{(2\pi)^{10}}\exp\ls-\pi\al'\tau_2\lp k-\frac{iLn}{\al'\tau_2}\rp^2-\frac{\pi L^2n^2}{\al'\tau_2}-\pi\al'\tau_2p^2\rs\lp\cdots\vphantom{\frac{1\pm e^{\pi i\tilde F}}{2}}\rp,\non\\
& = & iV_{10}Z_X^8\exp\lp-\frac{\pi L^2n^2}{\al'\tau_2}\rp\hlf\ls\lp Z_0^0\rp^4-\lp Z_1^0\rp^4-e^{\pi in}\lp Z_0^1\rp^4-e^{\pi in}\lp Z_1^1\rp^4\rs\non\\
& & \times\hlf\ls\lp\bar Z_0^0\rp^4-\lp\bar Z_1^0\rp^4-e^{\pi
in}\lp\bar Z_0^1\rp^4\mp e^{\pi in}\lp\bar Z_1^1\rp^4\rs. \eea The
upper (lower) sign  represents IIB (IIA) while $F$ and $\tilde F$
are the left- and right-moving world-sheet fermion numbers.

Using modular invariance, it is not difficult to show that in the
$w$-twisted sector, the partition sum with $g^n$ inserted is given
by \bea
{\mathcal Z}_w^n & = & \frac{iV_{10}}{4\pi^2\al'\tau_2}Z_X^8\exp\lp-\frac{\pi L^2\left|n-w\tau\right|^2}{\al'\tau_2}\rp\non\\
& & \times\hlf\ls\lp Z_0^0\rp^4-e^{\pi iw}\lp Z_1^0\rp^4-e^{\pi in}\lp Z_0^1\rp^4-e^{\pi i(n+w)}\lp Z_1^1\rp^4\rs\non\\
& & \times\hlf\ls\lp\bar Z_0^0\rp^4-e^{\pi iw}\lp\bar
Z_1^0\rp^4-e^{\pi in}\lp\bar Z_0^1\rp^4\mp e^{\pi i(n+w)}\lp\bar
Z_1^1\rp^4\rs. \eea To get the full partition sum in the $w$-twisted
sector we simply sum over $n$, \bea
Z_w & = & \frac{1}{|\Z|}\sum_n{\mathcal Z}_w^n=\frac{iLV_9}{2\pi\al'\tau_2}Z_X^8\sum_n\exp\lp-\frac{\pi L^2\left|n-w\tau\right|^2}{\al'\tau_2}\rp\hlf\ls\cdots\rs\hlf\ls\cdots\rs\non\\
& = & \frac{i}{2\pi\sqrt{\al'\tau_2}}V_9Z_X^8\left\{\vphantom{\times\sum_n\exp\lp-\frac{\pi\al'\tau_2(n-\hlf)^2}{L^2}-\frac{\pi\tau_2
L^2w^2}{\al'}-2\pi i\tau_1(n-\hlf)w\rp}\right.\non\\
& & \left.\frac{1}{4}\lp\left|\lp Z_0^0\rp^4-e^{\pi iw}\lp Z_1^0\rp^4\right|^2+\ls\lp Z_0^1\rp^4+e^{\pi iw}\lp Z_1^1\rp^4\rs\ls\lp\bar Z_0^1\rp^4\pm e^{\pi iw}\lp\bar Z_1^1\rp^4\rs\rp\right.\non\\
& & \qquad\qquad\left.\times\sum_n\exp\lp-\frac{\pi\al'\tau_2n^2}{L^2}-\frac{\pi\tau_2 L^2w^2}{\al'}-2\pi i\tau_1nw\rp\right.\non\\
& & \left.-\frac{1}{4}\lp\ls\lp Z_0^0\rp^4-e^{\pi iw}\lp Z_1^0\rp^4\rs\ls\lp\bar Z_0^1\rp^4\pm e^{\pi iw}\lp\bar Z_1^1\rp^4\rs\right.\right.\non\\
& & \qquad\left.\left.+\ls\lp Z_0^1\rp^4+e^{\pi iw}\lp Z_1^1\rp^4\rs\ls\lp\bar Z_0^0\rp^4-e^{\pi iw}\lp\bar Z_1^0\rp^4\rs\rp\right.\non\\
& &
\qquad\qquad\left.\times\sum_n\exp\lp-\frac{\pi\al'\tau_2(n-\hlf)^2}{L^2}-\frac{\pi\tau_2
L^2w^2}{\al'}-2\pi i\tau_1(n-\hlf)w\rp\right\},
\label{w-sector-partition-function} \eea where we have used the
usual interpretation $V_\R/|\Z|=2\pi L$ and where we have performed
a Poisson resummation.

From this expression, we can now read the key features.  The GSO
projection only agrees with the usual one in even twisted sectors.
In general we have $\exp(\pi iF)=\exp(\pi i\tilde F)=\exp(\pi iw)$
for type IIB, with the usual right-moving fermion flip for the type
IIA case. Explicitly, denoting periodicities and world-sheet fermion
number, we have sectors \be
\lp{\mathrm{NS}}+,{\mathrm{NS}}+\rp\qquad\lp{\mathrm{R}}+,{\mathrm{NS}}+\rp\qquad\lp{\mathrm{NS}}+,{\mathrm{R}}\pm\rp\qquad\lp{\mathrm{R}}+,{\mathrm{R}}\pm\rp,
\ee for $w$ even and \be
\lp{\mathrm{NS}}-,{\mathrm{NS}}-\rp\qquad\lp{\mathrm{R}}-,{\mathrm{NS}}-\rp\qquad\lp{\mathrm{NS}}-,{\mathrm{R}}\mp\rp\qquad\lp{\mathrm{R}}-,{\mathrm{R}}\mp\rp,
\ee for $w$ odd.

The momentum is integrally moded in the NS-NS and R-R sectors. For
space-time fermions in R-NS or NS-R, it is half-integrally moded.
Level matching leads to the criterion $N_L=N_R+nw$ for the bosons
and $N_L=N_R+(n-\hlf)w$ for the fermions, where $N_L$ and $N_R$
include zero point energies and the states satisfy the GSO
projection.  Subject to these constraints, the nine-dimensional mass
squared is given by \be \label{RohmB}
m^2=\frac{n^2}{(L)^2}+\frac{w^2(L)^2}{(\al')^2}+\frac{2}{\al'}\lp
N_L+N_R\rp, \ee for spacetime bosons and \be \label{RohmF}
m^2=\frac{(n-\hlf)^2}{(L)^2}+\frac{w^2(L)^2}{(\al')^2}+\frac{2}{\al'}\lp
N_L+N_R\rp, \ee for spacetime fermions.

The only possible tachyons  come from the (NS$-$,NS$-$) sector with odd
winding and no oscillators.  The zero point energy for that sector
is $N_L=N_R=-\hlf$ and level matching then forces $n=0$.  The
remaining spectrum of possible tachyons is given by \be
\label{Rohmtach} m^2=\frac{\lp 2k+1\rp^2(L)^2-2\al'}{(\al')^2}. \ee
There are no tachyons for $L>\sqrt{2\al'}$, but as $L$ decreases,
more and more tachyons enter the spectrum, always with two-fold
degeneracy.  Each tachyon is a space-time scalar.

The massless sector for even $w$ contains only the bosonic part of
the corresponding type II spectrum.  The fermionic spectrum is
purely massive due to the momentum shift.  For odd $w$, there is a
massless (NS$-$,NS$-$) scalar if and only if $(L)^2={2\al'\over
(2k+1)^2}$, for some integer $k$.  The fermionic spectrum is purely
massive unless $(L)^2={\al'\over 2}$, in which case there are two
pairs of massless $\SO(8)$ spinors, from the (R$-$,NS$-$) sector with
$n=0$, $w=-1$ or $n=1$, $w=1$, and from the (NS$-$,R$\mp$) sector with
$n=0$, $w=1$ or $n=1$, $w=-1$.  The (R$-$,R$\mp$) sector is purely
massive.

It is worth briefly commenting on two limits.  First, taking $L$ to
infinity decouples all winding sectors and the momenta become
continuous (we can no longer tell that bosons and fermions have
different moding). We are left with the standard spectrum of type
II. In the limit of $L$ going to zero, things are more interesting.
States with non-zero momentum decouple; in particular, all the
space-time fermions decouple.  The winding contributions become
continuous and we are left with the spectrum \be
\label{typezerosectors}
\lp{\mathrm{NS}}+,{\mathrm{NS}}+\rp\qquad\lp{\mathrm{NS}}-,{\mathrm{NS}}-\rp\qquad\lp{\mathrm{R}}+,{\mathrm{R}}\pm\rp\qquad\lp{\mathrm{R}}-,{\mathrm{R}}\mp\rp,
\ee which is precisely the spectrum of type 0B (0A), which includes
a continuous family of scalar tachyons.

\subsection{Closed strings on the T-dual of the Rohm circle}

One can also consider the T-dual of the previous configuration.
T-duality simply interchanges momentum and winding, so we can read
off the spectrum from our results in the previous subsection.  In
particular, let \be L'=\frac{\al'}{L}. \ee Then our states are again
classified by a pair of integers $n$ and $w$, and the GSO projection
will be given by $\exp(\pi iF)=\exp(\pi i\tilde F)=\exp(\pi in)$ for
IIB, and the corresponding sign-flip for IIA. In other words, for
$n$ even we will have sectors \be
\lp{\mathrm{NS}}+,{\mathrm{NS}}+\rp\qquad\lp{\mathrm{R}}+,{\mathrm{NS}}+\rp\qquad\lp{\mathrm{NS}}+,{\mathrm{R}}\pm\rp\qquad\lp{\mathrm{R}}+,{\mathrm{R}}\pm\rp,
\ee while for $n$ odd we will have \be
\lp{\mathrm{NS}}-,{\mathrm{NS}}-\rp\qquad\lp{\mathrm{R}}-,{\mathrm{NS}}-\rp\qquad\lp{\mathrm{NS}}-,{\mathrm{R}}\mp\rp\qquad\lp{\mathrm{R}}-,{\mathrm{R}}\mp\rp.
\ee

For bosonic states we have a nine-dimensional  mass squared \be
\label{RohmTdualB} m^2=\frac{n^2}{(L')^2}+\frac{w^2
(L')^2}{(\al')^2}+\frac{2}{\al'}\lp N_L+N_R\rp, \ee and level
matching condition $N_L=N_R+nw$, while for space-time fermions we
have \be \label{RohmTdualF} m^2=\frac{n^2}{(L')^2}+\frac{\lp
w-\hlf\rp^2 (L')^2}{(\al')^2}+\frac{2}{\al'}\lp N_L+N_R\rp, \ee and
level matching $N_L=N_R+n(w-\hlf)$.

Once again the only possible tachyons are scalars from the (NS$-$,NS$-$)
sector with no oscillators, $w=0$, and with $n=2k+1$. Then we have
\be \label{RohmTdualtach} m^2=\frac{\lp
2k+1\rp^2}{(L')^2}-\frac{2}{\al'}. \ee Thus there are no tachyons
for $L'<\sqrt{\al'/2}$, but as $L'$ increases, more and more
tachyons enter the spectrum.

For even $n$, the massless bosonic  spectrum is identical to the
corresponding type II on a circle, while the fermionic spectrum is
purely massive.  For odd $n$, the (NS$-$,NS$-$) sector can contain
tachyonic or massless states, as described above. The (R$-$,R$\mp$)
sector is purely massive, and the fermions are all massive except if
$(L')^2=2\al'$ for which there are two pairs of massless spinors.

In the limit of large $L'$, many states decouple and we have a
continuous family of tachyonic scalars. The spectrum is that of a
type 0 theory in flat space.  In the limit where $L'$ goes to zero,
only the states with $n=0$ retain a finite mass, and we recover the
spectrum of a type II theory in flat space.  All of these properties
follow simply by T-duality.

One can also construct this theory  more directly as a
$\Z_2$-orbifold of type II string theory on a circle.  Explicitly,
start with type II on a circle of radius $L'/2$.  The states of this
theory can again be described by a pair of integers, $n'$ and $w'$.
We then quotient this theory by the $\Z_2$ symmetry group generated
by $e^{\pi iw'}(-1)^{F_s}$.  This asymmetric orbifold theory then
has precisely the spectrum of the previous subsection. The untwisted
sector of the orbifold gives rise to the states with even $n$, while
the twisted sector furnishes the states with odd $n$.

\subsection{A heterotic example}

A tachyon-free non-supersymmetric heterotic string background was
constructed by Dixon and Harvey\cite{Dixon:1986iz}. It can be
obtained as an orbifold of the supersymmetric $E_8\times E_8$ theory
by $(-1)^{F_s}$, together with a shift of the $E_8\times E_8$ torus
by a half period along the lattice vector \be \delta = ((1/2)^8;0^8)
\ee in the integer weight lattice of $O(32)$.  In the fermionic
construction of $O(32)$, this amounts to a shift by one of the
fermion numbers $F$ and $F'$ for two sets of 16 fermions $\chi_I$,
$\chi'_I$ that realize the group symmetry.  In the twisted sector,
the lowest states carry two excitations above the vacuum, one from
each set of fermions; this lifts the ground state energy up to zero.
The massless spectrum thus consists of the bosonic part of the
supergravity multiplet, gauge bosons for $O(16)\times O(16)$, and
fermions of one chirality in the $(128,1)\oplus(1,128)$, and
fermions of the opposite chirality in the $(16,16)$. There is no
tachyon in the spectrum.

\subsection{Localized closed string tachyons}

Rotational orbifolds that break supersymmetry  can lead to closed string tachyons localized at the orbifold fixed locus.  The condensation of these tachyons has been argued to lead to a dynamic relaxation of the deficit angle of the orbifold, a decay to other localized geometries
~\cite{Adams:2001sv,Harvey:2001wm}.   Consider $\Cc/\Z_k$, parametrized by a complex
worldsheet scalar field $X$ with the group action $X\to\omega X$, $\omega=e^{\frac{2\pi i}{k}}$.
Without the chiral GSO projection, one has type 0 string theory on a nonsupersymmetric orbifold,
with tachyon masses
\be
\label{localtachmass}
\frac{\ap}{4} m_j^2 = -\frac12(1-\frac jk)
\ee
in the $j^{\rm th}$ twisted sector.  The untwisted sector $j=0$ houses the bulk type 0 tachyon, the twisted sectors with $j=1\dots k-1$ yield tachyons localized at the orbifold fixed point.  A type II version arises upon application of the chiral GSO projection, which requires $k$ to be odd~\cite{Adams:2001sv};
in this case the twisted sector ground states have masses
given by~\pref{localtachmass} for $j$ odd, while for $j$ even the masses are
\be
\label{typeIIlocaltachmass}
\frac{\ap}{4} m_j^2 = -\frac12(\frac jk)\ .
\ee
In particular, the untwisted sector no longer contains a bulk tachyon.

%%%%%%%%%%%%%%%%%%%%%%%%%%%%%%%%%%%%%%%%%%
%%%%%%%%%%%%%%%%%%%%%%%%%%%%%%%%%%%%%%%%%%

\subsection{The effective potential at one string loop}
\label{effectivepotential}

To leading (classical) order in the string  coupling, type IIA
string theory on the Rohm circle is a valid solution.  However, as
we have seen in equation (\ref{w-sector-partition-function}), at
one-loop there is a non-vanishing vacuum amplitude.  This amplitude
is expressed as an integral over the the usual
$\operatorname{PSL}(2,\Z)$ fundamental domain $\mathcal{F}$, \be Z =
\int_{\mathcal{F}}\frac{d^2\tau}{\tau_2}\sum_wZ_w(\tau), \ee and
should be added to the tree-level effective action which is the
usual supergravity action in flat space.  Since $Z$ depends on some
of the fields, particularly on the size $L$ of the Rohm circle,
tadpoles for these fields will be induced\footnote{In string frame,
$Z$ depends on $L$ and on the space-time metric through
the volume factor, but there is no dilaton tadpole generated
directly. Conversely, in Einstein frame, $Z$ is accompanied by a
factor $e^{2\Phi}$ which directly generates a tadpole for the
dilaton.} and the classical solution is corrected at order $g_s^2$.

Let us make an ansatz for the corrected solution by writing, \be
ds^2=-dt^2+e^{2\al(t)}ds_{8-n}^2+e^{2\rho(t)}ds_n^2+e^{2\lambda(t)}dx^2,\qquad
e^{\Phi(t)}, \ee with the other fields vanishing.  Here we are
imagining $n$ compact coordinates of radius $e^\rho$, $8-n$
non-compact coordinates with scale factor $e^\alpha$, and we have
defined $e^\lambda=L$.  Now although the partition function was
derived under the assumption that the metric and string coupling
were constants, we can still obtain the leading-order correction to
the action by adding it to the tree-level result, \be
\label{corrected-action} S=-\frac{1}{2\k_{10}^2}\int
d^{10}x\sqrt{-g}e^{-2\Phi}\ls R+4\p_\m\Phi\p^\m\Phi\rs-Z. \ee Here
$Z$ will be the expression (\ref{w-sector-partition-function}), but
to account for the extra $n$ compact directions, we must replace $n$
of the $Z_X(\tau)$ factors with \bea
V_\R Z_X &=& V_\R\lp 4\pi^2\al'\tau_2\rp^{-1/2}\left|\eta\right|^{-2}\non\\
&\longrightarrow&\left|\eta\right|^{-2}\sum_{N,W}\exp\ls-\pi\tau_2\lp
e^{-2\rho}\al'N^2+e^{2\rho}\frac{W^2}{\al'}\rp+2\pi
i\tau_1NW\rs\equiv Z_\rho. \eea The total one-loop partition
function  $Z$ then depends on $\lambda$, $\rho$, and $\alpha$
(through the remaining volume factors $V_{9-n}=\int
d^{9-n}x\exp[(8-n)\al]$).  We would like to focus on the region
where $L^2\gg\al'$.  We see immediately that every term with $w\ne
0$ will be exponentially suppressed in $L$, so we are left with \bea
Z &\approx& \int_{\mathcal{F}}\frac{d^2\tau}{\tau_2}Z_0,\non\\
&=& \int_{\mathcal{F}}\frac{d^2\tau}{\tau_2}\frac{1}{4\pi\sqrt{\al'\tau_2}}\lp\int d^{9-n}x\, e^{(8-n)\al}\rp Z_X^{8-n} Z_\rho^n\left|Z_1^0\right|^8\\
&& \qquad\times\sum_n\lp\exp\ls-\frac{\pi\al'\tau_2n^2}{L^2}\rs-\exp\ls-\frac{\pi\al'\tau_2(n-\hlf)^2}{L^2}\rs\rp,\non\\
&=& \int \frac{d^{10}x}{2(2\pi)^{n+2}}\int_{\mathcal{F}}\frac{d^2\tau}{\tau_2}e^{(8-n)\al}Z_X^{8-n}Z_\rho^n\left|Z_1^0\right|^8\non\\
&& \qquad\times\frac{L}{\al'\tau_2}\sum_m\lp\exp\ls-\frac{\pi
L^2m^2}{\al'\tau_2}\rs-\exp\ls-\frac{\pi L^2m^2}{\al'\tau_2}-\pi
im\rs\rp.\non \eea We have performed a Poisson resummation to  get
the final line.

From this final expression we can see that the integrand of the
modular integral is exponentially suppressed in $L$ except for the
region where $\tau_2\gtrsim L^2/\al'$.  In this region,
$|\eta|^{-16}|Z_1^0|^8\sim 1$, up to exponentially suppressed
corrections.  Also, if we further assume that $L\gg e^{\rho}$ and
$L\gg \al' e^{-\rho}$ (that is that the Rohm circle is much bigger
than the compact circles or their T-duals) then in this region of
large $\tau_2$, only the $N=W=0$ term contributes to $Z_\rho$ so
$Z_\rho\sim|\eta|^{-2}$.

Putting in these simplifications, we are left with
\bea
\label{sugraeffpotl}
Z &\sim& \int\frac{d^{10}x}{(2\pi)^{10}}\int^\infty\frac{d\tau_2}{\tau_2}e^{(8-n)\al}L\lp\al'\tau_2\rp^{\frac{n}{2}-5}\sum_{m\ \mathrm{odd}}\exp\ls-\frac{\pi L^2m^2}{\al'\tau_2}\rs,\non\\
&\sim& \chi_n\int d^{10}x\, e^{(8-n)\al+(n-9)\lambda},
\eea
where
\bea
\chi_n &=& (2\pi)^{-10}\lp\sum_{m\ \mathrm{odd}}|m|^{n-10}\rp\lp\int^\infty dx\,e^{-1/x}\, x^{\frac{n}{2}-6}\rp,\non\\
&=& 2(2\pi)^{-10}\lp 1-2^{n-10}\rp\zeta(10-n)\times\left\{\begin{array}{ll}\lp 4-\frac{n}{2}\rp !, & n\ \mathrm{even}, \\ 2^{\frac{n-9}{2}}\sqrt\pi\lp 8-n\rp !!, & n\ \mathrm{odd}.\end{array}\right.
\eea

Since this leading  contribution comes from the region of large
$\tau_2$, we should be able to interpret it as a field theory
effect, rather than being stringy in origin.  Indeed, this effect is
simply the usual field theoretic Casimir energy.  Note in particular
that the expression we found is independent of the string scale
$\al'$.  It is also independent of the compactification scale
$e^\rho$, as it should be since the low energy effective field
theory is insensitive to these scales in the large $L$ regime that
we have assumed.

For the T-dual of the Rohm compactification, one performs
essentially the same exercise as above, except that the light modes
are now the winding modes and the twisted circle radius is T-dual to
the one used above. The effective potential \pref{sugraeffpotl} is
simply modified by the substitution $\lambda\to-\lambda$.

We can now attempt to find a classical cosmological solution to the
corrected action. Following~\cite{Tseytlin:1991xk}, we define 
\be
\varphi=2\Phi-\lp 8-n\rp\al-n\rho-\lambda. \ee 
Then we can write our
action as \be S=\int d^{10}x\sqrt{-G_{00}}\ls
e^{-\varphi}G^{00}\lp(8-n)\dot\al^2+n\dot\rho^2+\dot\lambda^2-\dot\varphi^2\rp-\chi_ne^{(8-n)\al+(n-9)\lambda}\rs.
\ee We have introduced the metric component $G_{00}$ here in order
to obtain the equation from its variation, but we will set
$G_{00}=-1$ in the equations of motion.  We have also absorbed the
gravitational coupling into the definition of $\varphi$.

\begin{figure}[ht]
\label{SUGRAplot}
\begin{center}
\[
\mbox{\begin{picture}(230,220)(0,30)
\includegraphics[scale=.8]{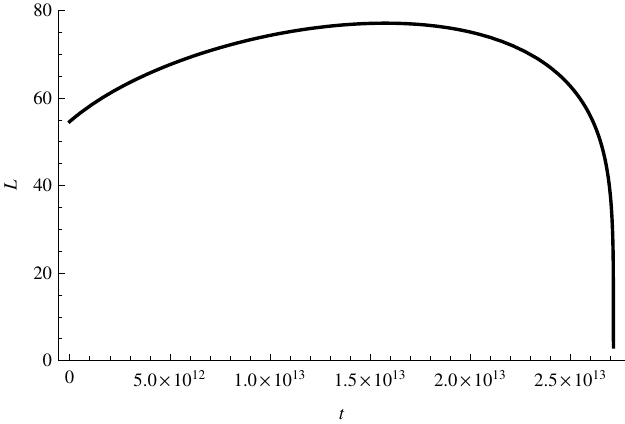}
\end{picture}}
\]
\vskip 0.2 in \caption{\it Behavior of the circle size, $L(t)$, as a
function of time.  This plot is for  the case of no transverse
circles, $n=0$, and initial values $\lambda(0)=4$, $\varphi(0)=-10$,
$\al(0)=0$, with $\dot\varphi(0)=\dot\al(0)=0$, $\dot\lambda(0)>0$
fixed by~(\ref{eq:SUGRAconstraint}).  Note that in the regime where
$L$ has fallen back below its initial value, the solution is very
well described by the attractor solution~(\ref{eq:attractor}).}
\end{center}
\end{figure}

The equations obtained from this action are
\bea
-\lp 8-n\rp\dot\al^2-n\dot\rho^2-\dot\lambda^2+\dot\varphi^2 &=& -\chi_ne^{\varphi+(8-n)\al+(n-9)\lambda},\label{eq:SUGRAconstraint}\\
\ddot\al-\dot\varphi\dot\al &=& \hlf\chi_ne^{\varphi+(8-n)\al+(n-9)\lambda},\\
\ddot\rho-\dot\varphi\dot\rho &=& 0,\\
\ddot\lambda-\dot\varphi\dot\lambda &=& \hlf\lp n-9\rp\chi_ne^{\varphi+(8-n)\al+(n-9)\lambda},\\
\ddot\varphi-\lp 8-n\rp\dot\al^2-n\dot\rho^2-\dot\lambda^2 &=&
-\hlf\chi_ne^{\varphi+(8-n)\al+(n-9)\lambda}. \eea 
If we set
$\dot\rho=0$ as an initial condition then it remains true through
out the evolution, and $\rho$ simply drops out of the remaining
equations. In this case, the equations above admit an exact solution
of the form \bea
\lambda(t) &=& \lambda_0+\frac{2}{10-n}\ln\lp t_c-t\rp,\non\\
\al(t) &=& \al_0-\frac{2}{(10-n)(9-n)}\ln\lp t_c-t\rp,\label{eq:attractor}\\
\varphi(t) &=& \varphi_0-\frac{2}{(10-n)(9-n)}\ln\lp t_c-t\rp,\non
\eea where the constant terms must satisfy \be \chi_n
e^{\varphi_0+(8-n)\al_0+(n-9)\lambda_0}=\frac{4(11-n)(8-n)}{(10-n)(9-n)},
\ee and where $t_c$ is a critical time at which the solution becomes
singular.  This is certainly not the most general solution for
arbitrary initial conditions, but it seems that if the system starts
at relatively large circle size and relatively weak string coupling
then it will eventually converge on this solution, though it can
take a significant amount of time.  An example of the evolution of
$L(t)=e^{\lambda(t)}$ is plotted in Figure~1.  The value of $t_c$
depends on the choice of initial conditions but it tends to scale as
follows \be t_c\sim
e^{\hlf\lp\varphi(0)+(8-n)\al(0)+(n-9)\lambda(0)\rp}. \ee

The attractor solution (\ref{eq:attractor}) causes the circle to
shrink towards string scale as a power of $(t_c-t)$, \be
L(t)=e^{\lambda_0}\lp t_c-t\rp^{\frac{2}{10-n}}, \ee while also
causing the scale factor $e^\al$ and the effective dilaton $\varphi$
to grow.  Intriguingly, the ten-dimensional string coupling is
constant in this solution, 
\be
\Phi=\hlf\lp\varphi_0+(8-n)\al_0+\lambda_0+n\rho_0\rp. 
\ee

Note that this process is not directly  related to tachyon
condensation, nor is it related to any strong coupling effect.  This
is simply the response of the background to the Casimir energy,
which, as we have seen, is field theoretic in nature.  Of course,
once we begin to approach the tachyonic regime, the derivation of
these equations is no longer valid.

%The structure of the equations is that of a collection of scalars in an inverted exponential potential.  The size of the twisted circle is attracted toward the string scale, where the tachyon develops and condenses just as the bottom is dropping out of the potential for $\alpha$ and $\lambda$.  The dilaton is attracted toward strong coupling in the process.

For the non-supersymmetric heterotic example discussed in section
2.3, there is also a one-loop potential which was evaluated
in~\cite{Dixon:1986iz}. There it was argued that the one-loop
potential is positive unlike the Rohm compactification or the
bosonic string, causing the dilaton to roll out toward vanishing
string coupling.

%%%%%%%%%%%%%%%%%%%%%%%%%%%%%%%%%%%%%%%%%%
%%%%%%%%%%%%%%%%%%%%%%%%%%%%%%%%%%%%%%%%%%
%%%%%%%%%%%%%%%%%%%%%%%%%%%%%%%%%%%%%%%%%%

\section{The Non-Perturbative Description}
\subsection{Type 0A}

We will start by considering the light-cone description of M-theory
on $\Tt^p$ given by Matrix theory~\cite{Banks:1996vh}.
% (for a review, see~\cite{Banks:1999az}).
For simplicity, we consider a rectangular torus with sides of length
$L_i$. The only other scale is the eleven-dimensional Planck scale
$\ell_p$. One can think of Matrix theory as a description of the
theory compactified on a light-like circle \be X^- \sim X^- + 2\pi
\LR \ee with $N$ units of light-cone momentum excited, so that
$P^+={N \over R}$. Alternatively, from the perspective of the
AdS/CFT correspondence one may think of the theory as arising from a
high boost of dynamics along the sub-Planckian M-theory circle of
type IIA, such that the back reaction of the boost causes the
effective size of that cycle to expand to macroscopic
size~\cite{Horowitz:1997fr}. The scales for the theory describing
$\Tt^p$ are $(L_i, R, \ell_p)$.

%This background is equivalent to a decoupling limit of type IIA string theory in the presence of a large number $N$ of D0 branes.   While the eleventh dimension in perturbative type IIA string theory is much smaller than the Planck scale, effectively it becomes large in the decoupling limit; on the geometrical side of the gauge/gravity duality, low energy corresponds to going toward the horizon of the extremal D0 geometry (and thus redshifting); the backreaction of the D0 source is momentum, which exerts pressure on the geometry and causes its proper size to expand near the source.

The non-perturbative description of this background is given in
terms of $p+1$-dimensional maximally supersymmetric $U(N)$
Yang-Mills compactified on a torus $\widehat{\Tt}^p$. The sides of
$\widehat{\Tt}^p$ have length \be \label{matcircsize} \Sigma_i =
{\ell_p^3\over \LR L_i}\ . \ee The Yang-Mills coupling and torus
volume determine the ``size" of the light-cone circle 
\be
\label{matcoupling} \gym^2/V_\S = {R^3 \over \ell_p^6}\quad, \qquad
\qquad V_\S = \prod_i \Sigma_i\ . \ee 
The action describing this
system is obtained by starting with $N=1$ Yang-Mills in ten
dimensions and dimensionally reducing to $p+1$ dimensions: 
\bea
S_{YM} &= & \frac{1}{g^2_{\rm YM}} \int \, d^{p+1}x \,   \Tr\Big( -{1\over
4} F_{\mu\nu} F^{\mu\nu}  -{1\over 2} \sum_{i=1}^{9-p}\left( D_\mu
\phi^i \right)^2 + \left[\phi^i, \phi^j \right]^2 \cr 
&& \qquad\qquad\qquad\qquad\qquad + {i\over
2} \bar\psi \Dslash \psi  + {1\over 2} \bar \psi \gamma^i [ \phi^i,
\psi]\Big)\ . \eea 
The symmetries of the action are a $Spin(9-p)$
R-symmetry together with the $Spin(p,1)$ Lorentz symmetry.

%A direct attempt to derive the non-perturbative description of these
%non-supersymmetric backgrounds is subtle. We describe how to do this
%in section~\ref{spacetime}. For the moment, we will simply propose a
%dual description and examine the physics of the model.

We are interested in matching the gauge theory description to
perturbative string theory. This arises upon making one of the
$L_i$, call it $L_M$, much smaller than the Planck scale, \ie\
$\S_M$ is large~\cite{Dijkgraaf:1997vv}. The string scale in terms
of the gauge theory variables is set by \be \S_M = {\alpha'\over R}\
. \ee

Let us now impose anti-periodic (twisted) boundary conditions on the
gauge theory fermions along this large M-theory circle $\S_M$. This
is a mild breaking of supersymmetry by boundary conditions which
does not affect the ultraviolet properties of the gauge theory. It
will turn out that the dynamics of the twisted gauge theory is that
of type 0A string theory~\cite{Seiberg:1986by}.

 In Appendix~\ref{gaugepotential}, we compute the one-loop perturbative
potential on the moduli space. It decays very rapidly for separated
branes because of the mildness of the supersymmetry breaking. This
provides evidence for the existence of an approximate moduli space.
However, for the most part, we will be discussing the strongly
coupled gauge theory for which the relevance of just the one-loop
potential is unclear without a non-renormalization argument.

In the weak coupling limit and with large $\S_M$,  the effect of the
supersymmetry breaking twist is to lift the energies of fermionic
states relative to the bosons.  The Casimir energies of bosons and
fermions now differ, and no longer cancel.  The calculation is
effectively 1+1 dimensional and (at zeroth order in the gauge
coupling) the same as the calculation which gives the ground state
energy of the NS sector of the fermionic string. Each of the
$N^2$ modes of the gauge theory contribute to the total energy: 
\be E_\ym = -\frac{N^2}{\S_M} \ ,
\ee where we have assumed that the gauge symmetry is not
spontaneously broken by separating the D-strings on the moduli
space. The small excitations around this state do not look like
Matrix strings, and it seems unlikely that there are quasi-BPS
excitations whose properties reflect those of the strongly coupled
theory.

In the strong coupling limit and again large $\S_M$, one expects
that the gauge dynamics locally confine the off-diagonal
excitations of the gauge theory, and the system reduces to Matrix
string dynamics with twisted boundary conditions along the Matrix
string.  The eigenvalues of the gauge theory scalars specify the
location of the strands of the Matrix string in the transverse
space, and are permuted by an element $\sigma$ of the symmetric
group by the boundary conditions \be x_i(2\pi) = x_{\sigma(i)}(0) \
. \ee There are as many strings in the state  as the number of
cycles in the symmetric group word $\sigma$, and the longitudinal
momentum $P^+_i={n_i\over R}$ of each string is determined by the
length $n_i$ of its corresponding cycle.

To leading order in the strong coupling limit, the strands of the
Matrix string are free. The effect of the supersymmetry breaking
boundary condition is to twist the fermions along the Matrix string
by $(-1)^{n_i}$ for the $i^{\rm th}$ strand. Thus the strings with
$n_i$ even don't feel the twist and have zero Casimir energy, while
the strings with $n_i$ odd have a ground state energy
$-1/(n_i\S_M)$. The ground state energy of the state to leading
order in the strong coupling expansion is 
\be\label{Esym0a} 
E_\ym = - \sum_i \frac{(1-(-1)^{n_i})}{2 n_i \S_M} \ . 
\ee 
Since the SYM energy is
the light  cone momentum $P^-$ in the Matrix theory interpretation,
using the transcription~\C{matcircsize}\ and~\C{matcoupling}, we
find that the individual Matrix strings of odd winding along $\S_M$
obey the mass shell condition \be P^+ P^- = -{1\over\alpha'} \ee
just as one expects for the type 0A tachyon, while the even winding
strings are massless.

Because the even winding strings don't feel the supersymmetry
breaking boundary conditions, there are fermion zero modes which act
on a degenerate ground state representation; this representation
constitutes the polarizations of the graviton supermultiplet.  In
the odd winding sectors, the fermion zero modes are eliminated by
the boundary conditions, and the ground state is indeed a scalar.
The Matrix strings also carry a $\Z_2$ quantum number -- the
even-odd parity of the winding $n_i$ -- which can be identified with
the conserved chiral worldsheet R-parity $(-1)^{F_L}$ of the type 0A
theory.

The even holonomy sectors have  the same spectrum as the type II
theory and carry even parity under $(-1)^{F_L}$; the odd holonomy
sectors are odd under $(-1)^{F_L}$ and have a spectrum built as
oscillator excitations above the scalar tachyonic vacuum.  In
particular, there is a massless level achieved by acting with the
lowest left- and right-moving fermion creation operators on the
vacuum, which contains the extra Ramond gauge fields of the type 0
theory.

In the non-compact theory on $\Rr^{9,1}$, the above estimate 
of the spectrum can be made precise.  The dual gauge theory is
simply maximally supersymetric Yang-Mills on $\Ss^1$; anti-periodic 
boundary conditions for the gauge theory fermions means
working in the NS sector of the theory.  The gauge theory flows in the infrared 
to the symmetric orbifold $(\Rr^8)^N/S_N$; we are interested in the NS sector
of this orbifold.

Twisted sectors in this orbifold describe collections of matrix strings with 
momenta $P_i^+=\frac{n_i}{R}$, with $\sum_i n_i = N$.  There is a twist
operator for each conjugacy class of the symmetric group, thus the operators
are characterized by the collection $\{n_i\}$ and implement cyclic 
permutations of groups of $n_i$ copies of $\Rr^8$.  Each individual
cyclic twist 
\be
X_k(\sigma+2\pi)=X_{k+1}(\sigma),\qquad k=1,\ldots,n_i;~n_i+1\equiv1,
\ee
can be diagonalized into a product of standard twist operators on complex scalars
\be
X(\sigma+2\pi)=e^{2\pi i\nu}X(\sigma),\qquad \nu=0,\coeff1{n_i},\ldots,\coeff{n_i-1}{n_i}.
\ee
The dimension of the component twist operators is $4\times\frac12\nu(1-\nu)$ for bosonic twist operators.  Fermion twist operators for the twist
\be
\psi(\sigma+2\pi)=e^{2\pi i\nu}\psi(\sigma)
\ee
have dimension $4\times\frac12\nu^2$, but one must decide 
what is the appropriate range of $\nu$.  For the Ramond sector
of the orbifold, corresponding to the supersymmetric matrix string,
the twist $\nu=k/n_i$ for a boson is paired with a twist $\nu+\frac12=k/n_i$
for a fermion.  The total dimension for a cyclic twist is
\be
4\times\frac12\sum_{k=0}^{n-1}\frac{k}{n_i}\Bigl(1-\frac{k}{n_i}\Bigr) +
4\times\frac12\sum_{k=0}^{n-1}\Bigl(\frac{k}{n_i}-\frac12\Bigr)^2 = n_i/2.
\ee
Summing up over the collection of cycles yields a total dimension $\sum_i n_i/2 = N/2$;
all the Ramond ground states are degenerate.  The matrix string interpretation is
a collection of strings at rest, which indeed should be degenerate in the supersymmetric theory.

The Neveu-Schwarz sector counterparts to these twist operators are obtained by 
spectral flow, and have dimension
\be
4\times\frac12\sum_{k=0}^{n-1}\frac{k}{n_i}\Bigl(1-\frac{k}{n_i}\Bigr) +
4\times\frac12\sum_{k=0}^{n-1}\Bigl(\frac{k}{n_i}\Bigr)^2 = n_i-1,
\ee
with the total dimension for the collection of cycles being $N$ minus the number of strings.
This result disagrees with the result \pref{Esym0a}, where up to the overall constant shift by $N$,
the energies of matrix strings should be of order $1/n_i$ in order to have a sensible 
light-cone physics interpretation; instead the light-cone energies of these states is $O(1)$
not $O(1/n_i)$.  

However, these states are the NS sector version of BPS states
of the orbifold CFT; they carry a large $R$-charge proportional to their conformal dimension.
One can find more suitable states by looking for states in the same twist sector which have 
the minimum possible $R$-charge.  These states have dimension
\be
4\times\frac12\sum_{k=0}^{n-1}\frac{k}{n_i}\Bigl(1-\frac{k}{n_i}\Bigr) +
4\times\frac12\Bigl[\sum_{k=0}^{n/2-1}\Bigl(\frac{k}{n_i}\Bigr)^2+
\sum_{k=n/2}^{n-1}\Bigl(\frac{k}{n_i}-1\Bigr)^2\Bigr] = \frac{n_i}2
\ee
for $n_i$ even, while for $n_i$ odd they have dimension
\be
4\times\frac12\sum_{k=0}^{n-1}\frac{k}{n_i}\Bigl(1-\frac{k}{n_i}\Bigr) +
4\times\frac12\Bigl[\sum_{k=0}^{(n-1)/2}\Bigl(\frac{k}{n_i}\Bigr)^2+
\sum_{k=(n+1)/2}^{n-1}\Bigl(\frac{k}{n_i}-1\Bigr)^2\Bigr] = \frac12\Bigl(n_i-\frac1{n_i}\Bigr).
\ee
This spectrum exactly matches our result \pref{Esym0a}\ up to the overall fixed 
additive constant $N/2$.  This gives us confidence that the calculation \pref{Esym0a} 
and its generalizations below are correct, and that there is no dynamical effect that 
we have overlooked.%
\footnote{In the closely related $(T^4)^N/S_N$ orbifold dual to quantum gravity on
$AdS_3\times \Ss^3\times \Tt^4$, the maximally twisted state with $n_1=N$ is 
the extremal nonsupersymmetric BTZ black hole with zero $R$-charge, \ie\ 
no angular momentum on $\Ss^3$, or rather it becomes so upon perturbing 
in the moduli space away from the symmetric orbifold point into the geometrical regime.
Of course, in the matrix string orbifold, the twist operator which mediates string
interactions is irrelevant, and so there is no such geometrical interpretation.}
 
\subsection{The type IIA Rohm compactification}

 The spectrum~\C{RohmB}\ and~\C{RohmF}\ of the Rohm compactification
arises in the gauge theory if one combines the supersymmetry
breaking boundary condition along $\S_M$ with a shift along another
cycle.  Let the Rohm cycle be $\S_\rho$. The energy of the winding
tachyon~\pref{Rohmtach}, translated to gauge theory variables, is
\be \label{gaugeWtach} E_\ym =(2k+1)^2\; \frac{V_\S}{2n_i\gym^2
\S_\rho^2\S_M^2} - \frac{1}{ n_i \S_M} \ . \ee The second term is
again the Casimir energy of broken supersymmetry; the first term is
the energy of $2k+1$ units of magnetic flux in the gauge theory.
Similarly, the spectrum of the momentum tachyon~\pref{RohmTdualtach}
arises from electric flux in the gauge theory \be \label{gaugePtach}
E_\ym = (2k+1)^2\;\frac{\gym^2\S_\rho^2}{2n_i V_\S} - \frac{1}{ n_i
\S_M} \ . \ee 
The requirement that the odd length Matrix strings
carry magnetic flux can be incorporated by combining the
$(-1)^{F_s}$ twist along $\S_M$ with a shift of the $U(1)$ part of $A_\rho$ by a half
period.   The shift forces a spatial variation of this potential, 
\be A_\rho = \frac{x_M}{2\S_M}\One\ , \ee
which results in magnetic flux.  For even length cycles, the total shift of $A_\rho$
around the matrix string is by a period of the gauge field; twisting the boundary condition 
by an element of $SU(N)$ allows the magnetic flux to vanish for such strings.
For odd length cycles, the total shift of $A_\rho$ is not commensurate with the 
periodicity of the gauge field; the shift cannot be relaxed by a gauge transformation
in the boundary condition,
and such strings are forced to carry magnetic flux.

The relation of the winding tachyon to magnetic flux is quite natural -- magnetic flux in the gauge theory carries the same charge as membrane wrapping number in M-theory; a string is a membrane wrapped around the M-theory circle, and so magnetic flux along the $\S_M$-$\S_\rho$ plane is string winding along $L_\rho$.  Similarly electric flux along $\S_\rho$ translates into momentum along $L_\rho$; however there does not seem to be a natural boundary condition in the gauge theory that leads to the spectrum~\pref{gaugePtach} (the natural boundary condition should instead involve the potential of the electric-magnetic dual theory).

Curiously, the twist of the  gauge theory is a flipped version of
the group action suggested by Bergman and Gaberdiel as the orbifold
of M-theory which yields the Rohm background~\cite{Bergman:1999km}.
These authors suggested that the Rohm compactification is the
quotient of M-theory on $\Tt^2$ by a $(-1)^{F_s}$ twist along the
Rohm circle, combined with a half-shift along the M-theory circle.
Our proposal for the gauge theory involves a $(-1)^{F_s}$ twist
along the dual of the M-theory circle, combined with a half-period
shift of the gauge field $A_\rho$ along the dual of the M-theory circle.

The flip in boundary conditions is an artifact  of the particular
states that represent the tachyonic vacuum of the Matrix string,
where one cannot tell the difference between our proposed twist and
the Bergman-Gaberdiel version in the dual theory.  The $(-1)^{F_s}$
twist looks like it occurs along the Rohm circle on the Matrix
string, even though it's along the dual of the M-theory cycle
because of the Matrix string holonomy.

As one follows along the sequence of windings of gauge theory
eigenvalues, one is moving along the spatial coordinate $\sigma$ on
the string worldsheet, and $\sigma$ is the fraction of $P^+$ carried
by the string up to that point along the string's spatial
coordinate.  In an odd holonomy sector, by the time one gets once
around the string one has gone around the M-theory circle an odd
number of times; the gauge theory boundary conditions that lead to
an odd number of magnetic flux units dictate through the duality
chain that one has also gone around the Rohm circle an odd number of
times.

In the gauge theory the Rohm circle winding is a half-shift of $A_\rho$
as one goes around the M-theory circle; since $A_\rho$ T-dualizes to $X_\rho$,
this is the T-dual of a half shift along the Rohm circle.  Thus
the effective observer walks around the string and sees several
things -- a $P^+$ that goes like $N$, an odd number of windings
around the Rohm circle, and twisted fermion boundary conditions.  
But the twisted boundary conditions arise in the parent
theory from a different cycle than it would appear they are coming
from based on the low energy description.

\subsection{The non-supersymmetric heterotic string}

The heterotic string is the limit of M-theory on $\Ss^1/\Z_2$
as the circle size becomes much smaller than the Planck scale.
The Matrix theory description for the $O(32)$ theory~\cite{Banks:1997it,Lowe:1997sx,Rey:1997hj,Horava:1997ns}
employs the gauge theory on the type I D-string in 1+1 dimensions,
or the various type IA/IB constructions involving orientifold planes
that can be reached from it via T-duality.
The resulting theory is a $(0,8)$ supersymmetric $O(N)$ gauge theory,
with $(0,8)$ matter in the symmetric tensor.  The chiral gauge anomaly
in the spectrum is cancelled by the addition of 32 chiral fermion multiplets
$\chi_I$, $\chi'_I$, $i=1\ldots16$ in the vector representation of $O(N)$.
The orientifold projection that reduces the gauge group from $U(N)$ to $O(N)$
breaks the center-of-mass $U(1)$ down to a discrete $\Z_2$
which acts as the GSO projection~\cite{Polchinski:1995df}.

The supersymmetry breaking twist is again $(-1)^{F_s}$ along the cycle $\S_M$ in the gauge theory, combined with a shift in the fermion numbers of the $\chi_I$, $\chi'_I$.   The twist generates a tachyonic ground state at level $-1/2$ for the right-movers to go with the vacuum at level $-1$ for the left-movers.   The GSO projection $(-1)^{F_R}=(-1)^{F_L+F'_L}$,
together with level matching (which is imposed by the $S_N$ permutation symmetry
of the strands of the Matrix string~\cite{Dijkgraaf:1997vv}),
requires excitations of both left- and right-moving fermions;
the ground states are at zero energy, and there is no tachyon in the spectrum.
The even holonomy sectors contribute the bosonic part of the supergravity multiplet, the gauge
bosons for $O(16)\times O(16)$, and fermions in the $(128,1)\oplus(1,128)$;
the odd holonomy sectors contribute fermions in the $(16,16)$.

%involves an orbifold of the gauge theory by a
%$\Z_2$ that acts as
%\bea
%X_1(2\pi) &=& -\Omega X_1(0) \Omega \\
%X_i(2\pi) &=& \Omega X_i(0) \Omega\quad,\qquad i=2,\ldots,9 \\
%\psi(2\pi) &=& -\Omega(\Gamma_{01}\psi)\Omega \\
%A_0(2\pi) &=& -\Omega A_0 \Omega
%\eea
%where $\Omega M\Omega = M^T$ for matrices $M$ in the adjoint of $U(N)$.
%The resulting theory is a $(0,8)$ supersymmetric $O(N)$ gauge theory, with $(0,8)$ matter in the symmetric tensor.  The chiral gauge anomaly in the spectrum is cancelled by the addition of 32 chiral fermion multiplets $\chi_I$, $\chi'_I$, $i=1\ldots16$ in the vector representation of $O(N)$.
%Further compactification on $\Ss^1/\Z_2\times \Tt^d$ results in gauge theory on
%$\Ss^1\times \Tt^d/\Z_2$, so that the orbifold acts simultaneously on the parameter space and the gauge group.  The resulting theory has $U(N)$ symmetry in the bulk of the torus, which reduces to $O(N)$ at the fixed points; $32/2^d$ fermions live at the fixed points to cancel anomalies.

\subsection{Localized closed string tachyons}

Orbifolds yielding localized closed string tachyons arise in the
present construction via the imposition of suitable twisted boundary
conditions in the gauge theory.  Bulk closed string tachyons arise
with $(-1)^{F_s}$ twisted boundary conditions along the M-theory
circle, which can be considered to be a $2\pi$ rotation in some
transverse plane.  Twisting by a $2\pi/k$ rotation acting by the
rotation group element \be \exp\Bigl[2\pi i\frac{k+1}k J_{89}\Bigr],
\ee where $J_{89}$ is the generator of rotations in the $89$ plane,
leads to a $\Z_k$ orbifold.  The  ground state energy for a cycle of
length $n_i$ is \be E_\ym =
\begin{cases}
-\frac{k-j_i}{k n_i \S_M}\, ,\quad\qquad & j_i ~~ {\rm odd} \cr
 &\cr
-\frac{j_i}{k n_i \S_M}\, ,\quad\qquad & j_i ~~ {\rm even}
\end{cases}
\ee where $j_i=n_i ~{\rm mod}~ k$.  This spectrum agrees with that
of the perturbative string theory appearing in~\pref{localtachmass}\
and~\pref{typeIIlocaltachmass}\ in the limit $n_i\to\infty$.

\section{The effects of interactions}

The subleading orders in the strong coupling expansion incorporate the effects of string interactions.  In the regime of parameters where a tachyon is present, one expects it to condense and shift the vacuum.  How does this phenomenon appear in the gauge theory?

Matrix string theory is the gauge theory version of light cone
string field theory.  Naively the vacuum is uncorrected in light
cone gauge, and the field theory typically describes an {\it in}
state consisting of a finite number of excitations about this
vacuum.  Each excitation is a Matrix string carrying a finite
fraction $n_i\propto N$ of longitudinal momentum, and so the gauge
theory energy vanishes in the large $N$ limit.  Interactions are
implemented by a twist operator that rearranges the holonomy of the
gauge theory eigenvalues along $\S_M$~\cite{Dijkgraaf:1997vv}.

A Matrix string that splits off a short string of finite cycle length
$\{n_i\}\to\{k,n_i-k\}$ (with $k$ fixed in the large $N$ limit)
liberates a finite amount of energy in the gauge theory,
proportional to $-1/k$, which is deposited in the kinetic energy of
the decay products.   Thus the instability in the gauge theory is an
instability to the emission of negative energy short strings (the
tachyons) carrying $P^+\to 0$; the fact that objects in Matrix
theory carry a minimum of one unit of longitudinal momentum,
$P^+=1/R$, serves to regulate the infrared divergences associated to
the condensation process.  At strong gauge theory coupling, which
translates into weak string coupling, the rate of tachyon
condensation is suppressed by the cost of changing the holonomy --
the holonomy changing operator is an irrelevant operator of scaling
dimension 3~\cite{Dijkgraaf:1997vv}, and is heavily suppressed in
the strong coupling limit.

When the lowest winding/momentum mode along the Rohm circle  is not
tachyonic, the system is still unstable.  In this situation, the
instability arises from the quantum effective potential for the
dilaton and the radius of the Rohm circle, exhibited in
section~\ref{effectivepotential}. This negative potential drives the effective
dilaton to strong coupling and the circle radius toward the regime
where the tachyon instability develops.  The full quantum theory
remains a cosmology.

To see this cosmology develop in the Matrix theory  representation,
one wants to see the development of the instability.  This should
occur just as in perturbative string theory, through loop diagrams.
Matrix string Feynman diagrams naively look in the strong coupling
limit like string light cone perturbation theory, with winding along
$\S_M$ a discrete version of longitudinal momentum and the holonomy
reconnection operator representing the three string vertex.  The
correction to the energy of a graviton state is the leading order
correction to the ground state energy in a sector of even holonomy,
arising at second order in perturbation theory.   Absent
divergences, a leading correction that arises at second order is
indeed negative, so the graviton will become slightly ``tachyonic."
The exponential suppression 
\be {\exp}\left[-\frac{\pi m^2L_\rho^2}{\alpha'\tau_2}\right] \ee 
of~\pref{sugraeffpotl} can be understood in
perturbative string theory from the classical string action.  
A string of size of order the string scale must grow to a size 
$\sim mL_\rho$ and shrink back down, in a worldsheet time of order
$\tau_2$.  Then $\partial_t X_\rho\sim mL_\rho/\tau_2$ and the
action is
\be S_{\rm ws} =\frac{1}{4\pi\alpha'} \int dt d\sigma
[(\partial_tX)^2-(\partial_\sigma X)^2]
    \sim \frac{(mL_\rho)^2}{\alpha' \tau_2} \ .
\ee 
The graviton polarizations most strongly  affected are those
that couple directly to the stress-energy of the winding along
$L_\rho$ (or in M-theory, membrane wrapping around $L_\rho$ and
$L_M$) -- the graviton $g_{\rho\rho}$ and the dilaton.

The gauge theory calculation has  the same structure at strong
coupling.   A state in the even holonomy sector splits into two odd
holonomy states, which is the leading order mechanism for
supersymmetry breaking to be communicated to this sector.  The odd
sector states carry magnetic flux, which appears and disappears in a
time $\tau_2$. The kinetic energy of the gauge field which must grow
from near zero to a large value and shrink back down is of the same
order of magnitude as the above estimate of the string result.

The DLCQ description provided by  the gauge theory regulated
infrared divergences by forcing all objects to carry a small nonzero
longitudinal momentum $1/R$.  These infrared divergences will make
their appearance when one tries to take the large $N$ limit
$N\to\infty$, $R\to\infty$, $N/R$ fixed.  One expects that the cloud
of infrared gravitons are trying to generate a large scale
modification of the space-time geometry, in particular the rolling
of the dilaton and Rohm circle radius that one sees from string
perturbation theory in space-time.  The gauge theory is not required
to have a time-dependent radius of the Rohm circle $\S_\rho$ at
finite $N$, but it may be forced to have time-dependence consistent
with the space-time geometry in order to cancel divergences
appearing in the large $N$ limit.

A measure of the developing condensate of the effective field theory is the expectation values of composite operators which realize the harmonic expansion
of the supergravity fields~\cite{Taylor:1999gq}.  These expectation values will be time-dependent
due to the decay of the initial state.  It would be interesting to see if at least qualitative properties of
the dual geometry could be calculated from the evolution of the state in the gauge theory.

\subsection{The final state}

The endpoint of the condensation process is a highly excited state in the gauge theory.  The emission of short strings liberates an amount of energy at least of order $N$.  Some properties of the state can be understood on the basis of thermodynamics.

The ground state of the system with the Rohm twist is intimately
connected to thermodynamics, since the antiperiodic boundary
conditions along the spatial Rohm circle amount to a double Wick
rotation of thermal boundary conditions.  In regimes where the gauge
theory has a geometric dual, one can study this ground state through
the dual geometry.  The regimes where geometry is valid can be
understood from the thermodynamic phase diagram of the gauge theory
on a torus~\cite{Martinec:1998ja,Martinec:1999sa} (for a review,
see~\cite{Martinec:1999bf}). For instance, consider the regime where
the dual unexcited ground state geometry is a generalization of the
AdS soliton~\cite{Horowitz:1998ha}, the double Wick rotation of the
black D$p$-brane geometry \bea \label{blackDpgeom}
ds^2 &=& H^{-1/2}\left(-f(r)dt^2+d\xx\cdot d\xx \right) + H^{1/2}\left( f^{-1}(r)dr^2 + r^2 d\Omega^2_{8-p}\right),\\
H(r) &=& \gym^2N\lstr^4/r^{7-p},\\
f(r) &=& 1-(r_0/r)^{7-p}. \eea The double Wick rotation of this
geometry sends the time coordinate $t$ to the coordinate $x_\rho$ of
the twisted Rohm circle, and one of the spatial coordinates $x$ to
the new time coordinate $t'$, which is then decompactified to yield
the generalized soliton geometry.  The free energy of the black
brane, divided by the temperature, is the action of the Euclidean
solution, which after the second Wick rotation is equal to $E_0
\Sigma_{t'}$ with $E_0$ the ground state energy.  Up to a numerical
factor of order one, \be E_0 \sim - (\gym^2)^{\frac{3-p}{5-p}}
N^{\frac{7-p}{5-p}} (\S_1\cdots\S_{p-2}\S_\rho)
    \S_M^{\frac{p-9}{5-p}} \ .
\ee

The black  D$p$-brane geometry is not always the dominant object
governing the thermodynamic ensemble.  In the parameter space of the
theory, other objects predominate, and it is the reinterpretation of
their free energy that provides the estimate of the ground state
energy with twisted boundary conditions.  In the strong coupling
regime, the dominant objects are the black M2-brane for $p=2$ \be
E_0 \sim - N^{3/2}\S_\rho/\S_M^2 \ee and the Hagedorn phase of the
Matrix string for $p=1$ \be E_0 \sim -N/\S_M \ . \ee There is a
crossover from D2 to M2 behavior at $\gym^2\sim N^{1/2}/\S_M$; and
similarly from D1 to Hagedorn at $\gym\sim N^{1/2}/\S_M$. At weak
coupling the ground state energy is simply the Casimir energy of
$N^2$ degrees of freedom on a twisted circle \be E_0 \sim
-N^2(\S_1\cdots\S_{p-2}\S_\rho) /\S_M^{p} \ . \ee This weak coupling
phase is not encountered in the Matrix theory application, since
Matrix strings only arise in a strong coupling limit.

The phase diagram of black  branes can thus be used to map out the
ground state energy of the twisted gauge theory in various parameter
regimes.  The example of 1+1 SYM on $\Ss^1$ is shown in Figure~2.

\begin{figure}[ht]
\begin{center}
\[
\mbox{\begin{picture}(210,260)(0,10)
\includegraphics[scale=.8]{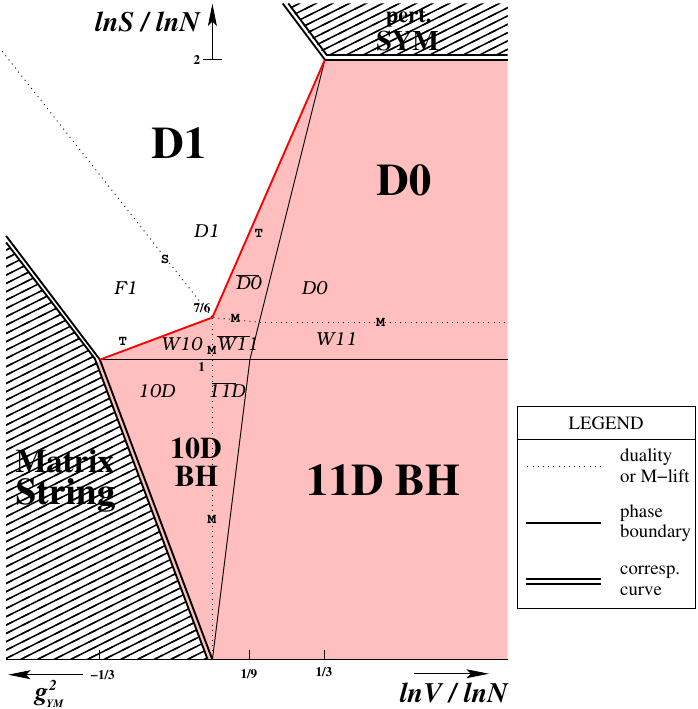}
\end{picture}}
\]
\caption{\it
Phase diagram for $p=1$.
}
\end{center}
\end{figure}

The pink shaded region describes black geometries that are localized on the cycle $\Sigma_{t'}$, and are thus are time-dependent geometries in the Rohm reinterpretation, {\it i.e.} not suitable for interpretation as the ground state of the twisted gauge theory.  However the regime related to matrix theory is the strong coupling regime, where the low energy physics for $p=1$ is that of a matrix string phase; the ground state energy is extensive along $\Sigma_{t'}$, negative and proportional to $N$.

The initial state of the system is that of a few long strings, characterized in the strongly coupled gauge theory by a word in the gauge theory of a fixed finite number of cycles in the large $N$ limit.  This state has an energy tending to zero at large $N$.  The ground state of the system has energy scaling like $-N^\alpha$ for $\alpha\ge 1$.  Thus, there is an amount of energy of this order released by the decay of the initial state via the emission of short strings of small $P^+$.  The gauge theory thermalizes (unless the constituents fly apart on the moduli space), much as one sees in gauge theory descriptions of black holes, and other Matrix theory cosmologies that have been investigated%
~\cite{
Li:2005ti,
Craps:2005wd,
Craps:2006xq,
Das:2006dr,
Das:2006dz,
Lin:2006ie,
Martinec:2006ak
}.

%%%%%%%%%%%%%%%%%%%%%%%%%%%%%%%%%%%%%%%%%%

\subsection{T-dual variants}
\label{T-dual}

As mentioned in section~\ref{perturbativestuff}, there are four
variants of the twisted circle compactification in perturbative
string theory -- IIA/IIB with momentum/winding tachyon forming when
the twisted circle reaches a critical string scale size.  In
perturbative string theory, all are related by T-duality
transformations.  However, non-perturbatively, T-duality is not a
symmetry.  For instance, in Matrix theory on $\Tt^3$ interpreted as
string theory on $\Tt^2$, T-duality on both cycles of the string
theory torus is S-duality in the $3+1$-dimensional gauge
theory~\cite{Ganor:1996zk,Susskind:1996uh}. This is not expected to
be a symmetry when N=4 supersymmetry is broken. Thus at the
non-perturbative level, all four variants are in principle distinct
theories.

The naive T-duality relations are
\bea
\tilde L_i &=& \frac{\lstr^2}{L_i} = \frac{\lpl^3}{L_M\,L_i }\\
\tilde g_s &=& g_s \prod_i\frac{\lstr}{L_i} \eea Consider the case
of $\Tt^3$ with the three cycles being the M-theory circle $L_M$,
the Rohm circle $L_\rho$, and a spectator circle $L'$.  If we
T-dualize along both $L_\rho$ and $L'$ in order to stay in type IIA
string theory, but pass to the variant with a winding tachyon, then
the ``duality" transformation in the gauge theory variables is \be
\label{IIAwinding} \gym^2\to 1/\gym^2\quad,\qquad
\S_\rho\to\S'/\gym^2 \quad,\qquad \S'\to \S_\rho/\gym^2\ . \ee The
gauge theory becomes very strongly coupled.   Electric modes are
interchanged with magnetic modes by the S-duality of the gauge
theory, and indeed the winding tachyon spectrum~\pref{gaugeWtach} is
exchanged with the momentum tachyon spectrum~\pref{gaugePtach}.

If one T-dualizes only the spectator circle $\S'$, one finds the
change of parameters \be \label{IIBmom}
\gym^2\leftrightarrow\frac{\S_\rho}{\S'}\quad,\qquad
\S'\to\S_\rho/\gym^2\quad,\qquad\S_\rho\to\S_\rho \ee which does not
involve S-duality; if one T-dualizes instead just the Rohm circle,
the transformation is \be \label{IIBwinding}
\gym^2\leftrightarrow\frac{\S'}{\S_\rho}\quad,\qquad
\S_\rho\to\S'/\gym^2\quad,\qquad\S'\to\S' \ee which again
interchanges electric and magnetic modes  as expected. It should be
emphasized that none of the
transformations~\pref{IIAwinding},~\pref{IIBmom}\
or~\pref{IIBwinding}\ is an exact symmetry of the non-perturbative
theory; rather, the transformations indicate which regimes of
parameter space should be considered such that the low-energy
dynamics is best described as, for example, type IIB with a winding
tachyon.

There is a related approach to finding T-dual models that is well
worth exploring. Let us take the spectator circle $L'\rightarrow
\infty$ so we are left with a $2+1$-dimensional Yang-Mills theory.
Supersymmetric type IIB string theory in ten dimensions is described
by the membrane theory compactified on $\Tt^2$~\cite{Sethi:1997sw,
Banks:1996my}. This is the desired exchange of electric and magnetic
degrees of freedom in $2+1$ dimensions to implement T-duality.

There is a beautiful realization of the membrane theory via
Chern-Simons gauge theory with maximal supersymmetry for a small
number of membranes~\cite{Bagger:2007jr}, and an extension to
M2-branes on an orbifold good for any $N$ (but with less manifest
supersymmetry)~\cite{Aharony:2008ug}. Unfortunately but expectedly,
the theory of M2-branes is still in a non-perturbative regime.

It is then natural to consider these Chern-Simons superconformal
field theories on $\Tt^2$ with twisted boundary conditions along a
large cycle of the torus. The strong coupling limit of these models
should provide non-perturbative definitions of type 0B and twisted
type IIB compactifications. It would still, nevertheless, be
interesting to see how much one can learn about non-perturbative
string theory from these compactified Chern-Simons theories even in
the perturbative limit -- both with and without supersymmetry
breaking twists.

%%%%%%%%%%%%%%%%%%%%%%%%%%%%%%%%%%%%%%%%%%

\section{Comparison to previous work}

The orbifold action that reproduces the perturbative spectrum of the
Rohm compactification twists by $(-1)^{F_s}$ along the dual M-theory
circle $\S_M$, together with a half-period shift of the gauge
potential $A_M$ along the Rohm circle (for winding tachyons).  This
action is a flipped version of the orbifold action argued by Bergman
and Gaberdiel~\cite{Bergman:1999km} to be the orbifold of M theory
that descends to the Rohm compactification in the perturbative
limit.   The perturbative Rohm theory is obtained by twisting by
$(-1)^{F_s}$ along the {\it Rohm} circle; Bergman and Gaberdiel
argued that this twist should be accompanied by a shift of a half
period along the {\it M-theory} circle.  We have found the same
twist, except with the roles of the Rohm circle and the M-theory
circle interchanged for the
$(-1)^{F_s}$ action in the gauge theory.%
\footnote{If the twisted fermion boundary condition in the gauge
theory  is placed along the Rohm circle, the fermion spectrum is
lifted by an amount that is large in string units; there are no
worldsheet fermions in the matrix string effective action.}

Adopting the Bergman-Gaberdiel orbifold action on the space-time
fields, Banks and Motl~\cite{Banks:1999tr} followed standard
arguments for how such an orbifold group should act on the fields of
the gauge theory -- one asks that the fields  be invariant under the
orbifold action up to a gauge transformation.  The resulting model
is a $2+1$-dimensional quiver gauge theory based on the gauge group
$U(N) \times U(N)$  (where $N$ is the the number of units of DLCQ
light-cone momentum in the system); there is a twisted boundary
condition that exchanges the two $U(N)$ factors as one goes once
around the Rohm circle.  The matter consists of scalars in the
adjoint of $U(N)\times U(N)$ obeying these same boundary conditions,
and fermions in the bifundamental with appropriate boundary
conditions.

The key issue is that  the model possesses no supersymmetry in the
ultraviolet.  The two-loop potential between gravitons was found to
be ultra-violet divergent which would seem to invalidate any
reasonable space-time interpretation.  By standard UV-IR
arguments~\cite{Peet:1998wn}, an ultraviolet divergence in the gauge
theory is a symptom of an infrared issue in spacetime; one would
need to fine tune the asymptotic boundary conditions in spacetime in
order to find a sensible UV gauge theory.  The authors concluded
that this background either does not exist as a non-perturbative
solution of string theory or is perhaps flowing to a
non-supersymmetric AdS solution.

Another possibility  is that the standard Matrix approach does not
provide a correct non-perturbative description of the Rohm circle.
Our alternative proposal keeps the asymptotic maximal supersymmetry
of the gauge theory, and only softly breaks it in the infrared. Thus
the sorts of UV difficulties found in the standard approach are
avoided. It would be very interesting to find a derivation of this
proposal from a space-time decoupling procedure along the lines
of~\cite{Seiberg:1997ad}\ though there is no guarantee that such a
derivation exists.

%One of the subtleties is determining the correct T-duality in the
%presence of $(-1)^{F_s}$ twists. For example, we can realize the
%twist by the Melvin identification \be \tau: \, (x, z) \rightarrow
%(x+2\pi, z e^{i\beta}) \ee where $x$ is real and $z$ is complex. For
%us, $\beta=2\pi$ and the space-time metric is flat: \be ds^2 =
%(dx)^2 + |dz|^2. \ee
% Usually, we change coordinates by extracting an
%angle from $z$ by defining $z  = w e^{i(x+\phi)}. $ The metric
%expressed in terms of $(x,w,\phi)$ is more complex \be ds^2 = |dx|^2
%+ |dw + iw (dx + d\phi)|^2  \ee but $x$ and $\phi$ now have
%canonical identifications $x \sim x + 2\pi, \, \phi\sim \phi+2\pi$.
%We are now free to perform a T-duality on $x$ but the result will
%generate a non-trivial metric and $B$-field. There are actually many
%circles we could choose to T-dualize \be x = ax' + b\phi', \quad
%\phi = cx' + d\phi' \ee where $ad-bc=1$ and $(a,b,c,d)$ are
%integers. With this choice, the coordinates $(x', \phi')$ have
%period $2\pi$. For an appropriate choice of $x'$, the resulting
%metric is again flat but there is a $2\pi$ rotation in a transverse
%plane coupled with a torsion $B$-field. The gauge theory on a brane
%in such a background

It is also important to note that  properties of the weakly coupled
gauge theory cannot be used here -- the weakly coupled gauge theory
has no ``long string" states that are the basis of the initial
unstable state of the Rohm background.  The initial state only
exists as a metastable state in the strongly coupled gauge theory,
and naively there are no quasi-BPS states of the weakly coupled
theory whose properties carry over to the strongly coupled regime
dual to geometry.  Furthermore, it is not the ground state of the
strongly coupled theory that is dual to the initial metastable state
of the dual string theory; rather it is a delicately tuned state of
long holonomy, with energy a positive power of $N$ above the ground
state, whose unwinding is the gauge theory description of the vacuum
decay.

%%%%%%%%%%%%%%%%%%%%%%%%%%%%%%%%%%%%%%%%%%
%%%%%%%%%%%%%%%%%%%%%%%%%%%%%%%%%%%%%%%%%%

\section*{Acknowledgements}

S.~S. would like to thank the KITP, Santa Barbara for hospitality
during the conclusion of this project, and D.~R. would like to thank them for hospitality at an intermediate stage. The work of E.~M. is
supported in part by DOE grant DE-FG02-90ER-40560. The work of D.~R.
is supported in part by the National Science Foundation under Grant
No. PHY-0455649. The work S.~S. is supported in part by NSF Grant
No. PHY-0758029 and by NSF Grant No.~0529954.

\newpage
\appendix
\section{Perturbative Aspects of the Gauge Theory}
\label{gaugepotential}

In this Appendix, we will examine how the flat directions are
modified in supersymmetric Yang-Mills theory with broken
supersymmetry along the $\S_1$ direction. We will compute the
$1$-loop potential induced by the supersymmetry breaking boundary
conditions. The intuition we would like to confirm is that for a
sufficiently large $\S_1$, the extent of supersymmetry breaking is
very small and the physics should be well described by the
supersymmetric theory with approximate flat directions in the scalar
potential. The dynamics along these flat directions gives rise to
gravitons and the notion of space-time.

So let us consider giving an expectation value to some of the scalars
$\phi^i$. In terms of the $SO(9-p)$ invariant distance $b$ between
two eigenvalues of $\phi^i$, we will have W-bosons with mass \be
m_W^2 \sim b^2 \ee which we want to integrate out of the theory to
obtain an effective action for dynamics on the Coulomb branch. The
generic expectation value breaks the gauge symmetry to the Cartan
$U(1)^N$. These are the light degrees of freedom. The heavy
off-diagonal degrees of freedom have actions of the form \be
\label{coulact} S = {1\over \gym^2} \int d\tau d^{p}x \,  \left(-
\partial_\mu X \partial^\mu X + b^2 X^2 \right). \ee We can Fourier
expand each bosonic field $X$ in the spatial directions. For fixed
spatial momenta $n_i$, the quadratic fluctuation operator is then
given by \be H =  \partial_t^2  + {n_i^2 \over \Sigma_i^2} + b^2.
\ee We then need to sum over all $n_i$. The effective potential for
$b$ generated by integrating out $X$ is given by \be \exp \left( i
\int d\tau d^p x \, V_{\rm eff} \right) = {\det}^{{-1/2}} (H). \ee
It is simpler to evaluate the derivative of this quantity with
respect to $b^2$. Diagrammatically, this gives the propagator \be
\frac{\partial V_{\rm eff}}{\partial b^2} = \int d\tau d^px\langle
X^2(t,x)\rangle = \frac12  V_\Sigma \int d\omega\sum_{n_i} \frac
1{\omega^2 + (n_i/\Sigma_i)^2+b^2+i\epsilon} \ . \ee The sum over
$n_1$ may be performed, yielding \bea \label{bospfn} \frac{\partial
V^{\it (bos)}_{\rm eff}}{\partial b^2} &=&  \frac12 V_\Sigma \int
d\omega\sum_{n_i,\; i>1}
\frac{\pi\coth\left[\pi\Sigma_1\omega_\nn \right]}{\omega_\nn},\\
\omega_\nn &=&
\Bigl[\omega^2+\sum_{i>1}(\coeff{n_i}{\Sigma_i})^2+b^2+i\epsilon\Bigr]^{1/2}.
\eea Splitting this apart as, \be \frac{\coth[x]}x = \frac 1x +
\frac 1{2x}\frac 1{e^{2x}-1}, \ee we recognize the vacuum energy of
one dimension lower in the first term, and the thermal energy in one
lower dimension in the second term (or rather their derivatives with
respect to $b^2$).

Now in the supersymmetric case, there are equal and opposite
contributions to the effective potential from the bosons and
fermions, and the vacuum energy vanishes. In the twisted case,
however, we need to make the replacement \be n_1 \rightarrow n_1 +
{1\over 2} \ee in~\C{bospfn}\ to obtain the fermion partition
function.  The result is to replace $\coth[x]/x$ in~\C{bospfn}\ by
\be \frac{\tanh[x]}x = \frac 1x - \frac 1{2x}\frac 1{e^{2x}+1}. \ee
Combining the boson and fermion contributions, one finds \be
\frac{\partial V_{\rm eff}}{\partial b^2}  = 8\times\frac12 V_\Sigma
\int d\omega\sum_{n_i,\; i>1}
\frac{\pi}{\omega_\nn}\frac1{e^{4\pi\Sigma_1\omega_\nn}-1} \ee Note
that the contributions from the ultraviolet are exponentially
suppressed by the thermal distribution functions. The potential
decays very rapidly as a function of $b$ and $\S_1$. This is very
much in accord with the one-loop potentials computed in the
time-dependent models~\cite{Craps:2006xq,Martinec:2006ak}.

%%%%%%%%%%%%%%%%%%%%%%%%%%%%%%%%%%%%%%%%%%
%%%%%%%%%%%%%%%%%%%%%%%%%%%%%%%%%%%%%%%%%%

%%%%%%%%%%%%%%%%%%%%%%%%%%%%%%%%%%%%%%%%%%
%%%%%%%%%%%%%%%%%%%%%%%%%%%%%%%%%%%%%%%%%%

\newpage
%%%%%%%%%%%%%%%%%%%%%%%%%%%%%%%%%%%%%%%%%%%%%%%%%%%%%%%%%%%%

%\bibliographystyle{utphys}
%\bibliography{myrefs}
%\providecommand{\href}[2]{#2}\begingroup\raggedright\begin{thebibliography}{10}

\providecommand{\href}[2]{#2}\begingroup\raggedright\endgroup

\end{document}